\documentclass[12pt]{article}

\usepackage{amsfonts, amsmath}

\newcommand{\be}{\begin{equation}} \newcommand{\ee}{\end{equation}}

\begin{document}
\title{Generalized Uncertainty Relation in  Thermodynamics} \thispagestyle{empty}

\author{A.E.Shalyt-Margolin\thanks
{Phone (+375) 172 883438; e-mail
alexm@hep.by}\hspace{10pt} and \hspace{5pt} A.Ya.Tregubovich
\thanks{Phone (+375) 172 840441; e-mail a.tregub@open.by}
}
\date{}
\maketitle
 \vspace{-15pt}
{\footnotesize\noindent {\large $^*$} National Centre of High Energy and
Particle Physics Bogdanovich Str.153, Minsk \hspace*{8pt} 220040, Belarus\\
{\large $^\dagger$} Institute of Physics National Academy of Sciences
                   Skoryna av.68, Minsk\\\hspace*{8pt} 220072, Belarus}\\

{\bf\small\noindent Abstract}\\ {\footnotesize
 A generalization of the thermodynamic uncertainty relations is proposed.
 It is done by introducing of an additional term proportional to the interior energy
  into the standard thermodynamic uncertainty relation that leads to existence
  of the lower limit of inverse temperature.}
 \vspace{0.5cm}
{\ttfamily{\footnotesize
\\ PACS: 03.65; 05.70\\ \noindent Keywords:
                   generalized   uncertainty relations; generalized
                   uncertainty\\ relations in thermodynamics}}

\rm\normalsize \vspace{0.5cm}

 It is well known that in thermodynamics an inequality for the pair interior energy -
 inverse temperature, which is completely analogous to the standard uncertainty
 relation in quantum mechanics \cite{r1} can be written down \cite{r2} -- \cite{r4}. The
 only (but essential) difference of this inequality from the quantum mechanical
 one is that the main quadratic fluctuation is defined by means of
 classical partition function rather than by quantum mechanical expectation values.
 In the last 14 - 15 years a lot of papers appeared in which the  usual
 momentum-coordinate uncertainty relation has been modified at very high
 energies of order Planck energy $E_p$ \cite{r5}--\cite{r9}. In this note we
 propose simple reasons for modifying the thermodynamic uncertainty relation at
 Planck energies. This modification results in existence of the minimal
 possible main quadratic fluctuation of the inverse temperature. Of course we
 assume that all the thermodynamic quantities used are properly defined so that
 they have physical sense at such high energies.

We start with usual Heisenberg uncertainty relations \cite{r1} for momentum -
coordinate:
\begin{equation}\label{U1}
 \Delta x\geq\frac{\hbar}{\Delta p}.
\end{equation}
 It was shown that at the Planck
 scale a high-energy term must appear:
\begin{equation}\label{U2}
\Delta
x\geq\frac{\hbar}{\Delta p} + \, const\, L_{p}^2\,\,\frac{\Delta p}{\hbar}.
\end{equation}
where $L_{p}$ is the Planck length $L_{p}^2 = G\hbar /c^3 \simeq
1,6\;10^{-35}m$. In \cite{r5} this term is derived from the string
theory, in \cite{r6}
 it follows from the simple estimates of Newtonian gravity and quantum mechanics,
 in \cite{r7} it comes from the black hole physics, other methods can also be
 used \cite{r8},\cite{r9}. Particularly the coefficient $const$ is shown to be unity in
 paper \cite{r6}.
Relation (\ref{U2}) is quadratic in $\Delta p$
\begin{equation}\label{U4}
 L_{p}^2\, ({\Delta p})^2 - \hbar\,\Delta x\Delta p + \hbar^2 \leq0
\end{equation}
 and therefore leads to the fundamental length
\begin{equation}\label{U5}
 \Delta x_{min}=2L_{p}
\end{equation}
  Using relations (\ref{U2}) it is easy to obtain a similar relation for the
 energy - time pair. Indeed (\ref{U2}) gives
\begin{equation}\label{U6}
\frac{\Delta x}{c}\geq\frac{\hbar}{\Delta p c }+L_{p}^2\,\frac{\Delta
p}{c \hbar},
\end{equation}
then
\begin{equation}\label{U7}
\Delta t\geq\frac{\hbar}{\Delta
E}+\frac{L_{p}^2}{c^2}\,\frac{\Delta p c}{\hbar}=\frac{\hbar}{\Delta
E}+t_{p}^2\,\frac{\Delta E}{ \hbar}.
\end{equation}
where the smallness of $L_p$ is taken into account so that the difference
between $\Delta E$ and $\Delta (pc)$ can be neglected and $t_{p}$  is the
Planck time $t_{p}=L_p/c=\sqrt{G\hbar/c^5}\simeq 0,54\;10^{-43}sec$.
Inequality (\ref{U7}) gives analogously to (\ref{U2}) the lower boundary
for time $\Delta t\geq2t_{p}$ determining the fundamental time
\begin{equation}\label{U10}
 \Delta t_{min}=2t_{p}.
 \end{equation}
 Thus, the inequalities discussed can be rewritten in a standard form
\begin{equation}\label{U11}
\left\{ \begin{array}{ll}
\Delta x & \geq\frac{\displaystyle\hbar}{\displaystyle\Delta p}+
\left(\frac{\displaystyle\Delta p}{\displaystyle P_{pl}}\right)\,
\frac{\displaystyle\hbar}{\displaystyle P_{pl}} \\
 & \\
 \Delta t & \geq\frac{\displaystyle\hbar}{\displaystyle\Delta E}+
 \left(\frac{\displaystyle\Delta E}{\displaystyle E_{p}}\right)\,
 \frac{\displaystyle\hbar}{\displaystyle E_{p}}
\end{array} \right.
\end{equation}
where $P_{pl}=E_p/c=\sqrt{\hbar c^3/G}$.
 Now we
consider the thermodynamics uncertainty relations between the inverse temperature
and interior energy of a macroscopic ensemble
\begin{equation}\label{U12}
\Delta \frac{1}{T}\geq\frac{k}{\Delta U}.
\end{equation}
where $k$ is the Boltzmann constant. \\ N.Bohr \cite{r2} and
W.Heisenberg \cite{r3} first pointed out that such kind of
uncertainty principle should take place in thermodynamics. The
thermodynamic uncertainty  relations (\ref{U12})  were proved by
many authors and in various ways \cite{r4}. Therefore their
validity does not raise any doubts. Nevertheless, relation
(\ref{U12}) was proved in view of the standard model of the
infinite-capacity heat bath encompassing the ensemble. But it is
obvious from the above inequalities that at very high energies the
capacity of the heat bath can no longer to be assumed infinite at
the Planck scale. Indeed, the total energy of the pair heat bath -
ensemble may be arbitrary large but finite merely as the universe
is born at a finite energy. Hence the quantity that can be
interpreted as the temperature of the ensemble must have the upper
limit and so does its main quadratic deviation. In other words the
quantity $\Delta (1/T)$ must be bounded from below. But in this
case an additional term should be introduced into (\ref{U12})
\begin{equation}\label{U12a}
\Delta \frac{1}{T}\geq\frac{k}{\Delta U} + \eta\,\Delta U
\end{equation}
where $\eta$ is a coefficient. Dimension and symmetry reasons give
$$
\eta = \frac{k}{E_p^2}.
$$
As in the previous cases inequality (\ref{U12a}) leads to the fundamental
(inverse) temperature.
\begin{equation}\label{U15}
T_{max}=\frac{\hbar}{2t_{p}
k}=\frac{\hbar}{\Delta t_{min} k}, \quad
\beta_{min} = {1\over kT_{max}} =  \frac{\Delta t_{min}}{\hbar}
\end{equation}
Thus, we obtain the system of generalized uncertainty relations in a symmetric
form
\begin{equation}\label{U17}
\left\{
\begin{array}{lll}
\Delta x & \geq & \frac{\displaystyle\hbar}{\displaystyle\Delta p}+
\left(\frac{\displaystyle\Delta p}{\displaystyle P_{pl}}\right)\,
\frac{\displaystyle\hbar}{\displaystyle P_{pl}} \\
&  &  \\
\Delta t & \geq & \frac{\displaystyle\hbar}{\displaystyle\Delta E}+
\left(\frac{\displaystyle\Delta E}{\displaystyle E_{p}}\right)\,
\frac{\displaystyle\hbar}{\displaystyle E_{p}}\\
  &  &  \\
  \Delta \frac{\displaystyle 1}{\displaystyle T}& \geq &
  \frac{\displaystyle k}{\displaystyle\Delta U}+
  \left(\frac{\displaystyle\Delta U}{\displaystyle E_{p}}\right)\,
  \frac{\displaystyle k}{\displaystyle E_{p}}
\end{array} \right.
\end{equation}
or in the equivalent form
\begin{equation}\label{U18}
\left\{
\begin{array}{lll}
\Delta x & \geq & \frac{\displaystyle\hbar}{\displaystyle\Delta p}+
L_{p}^2\,\frac{\displaystyle\Delta p}{\displaystyle\hbar} \\
  &  &  \\
  \Delta t & \geq &  \frac{\displaystyle\hbar}{\displaystyle\Delta E}+
  t_{p}^2\,\frac{\displaystyle\Delta E}{\displaystyle\hbar} \\
  &  &  \\
  \Delta \frac{\displaystyle 1}{\displaystyle T} & \geq &
  \frac{\displaystyle k}{\displaystyle\Delta U}+
  \frac{\displaystyle 1}{\displaystyle T_{p}^2}\,
  \frac{\displaystyle\Delta U}{\displaystyle k}
\end{array} \right.
\end{equation}
Here $T_{p}$ is the Planck temperature: $T_{p}=\frac{E_{p}}{k}$.
\\ In the conclusion we would like to note that the restriction on
the heat bath made above turns the equilibric  partition function
to be non-Gibbsian \cite{r10}.
\\ After the issue of our principal
work \cite{r11} devoted to the unification of the generalized
uncertainty relations in quantum theory and thermodynamics, Carlos
Castro has published an electronic preprint \cite{r12}, where with
a reference to our work he is attempting at justification of our
primary result proceeding from more common considerations.
However, he resorts to substitution, and we are of the opinion
that:
\\
$$ t\mapsto\frac{1}{T}, \hbar\mapsto k,\triangle E\mapsto
\triangle U$$
\\
could not be accepted as a rigorous proof for the primary result
of our paper. There is a reason to believe that a rigorous
justification for the last (thermodynamic) inequalities in systems
(\ref{U17}) and (\ref{U18}) may be made by the way of the
deformation of Gibbs distribution.
\\
Let us outline the main aspects of above-considered deformation.
In our opinion it could be obtained as the result of
density-matrix deformation in Statistical Mechanics (see
\cite{r13}, Section 2, Paragraph 3):
\begin{equation}\label{U19}
\rho=\sum_{n}\omega_{n}|\varphi_{n}><\varphi_{n}|,
\end{equation}
where probability is given by
\\
$$\omega_{n}=\frac{1}{Q}\exp(-\beta E_{n}).$$
\\
Deformation of density matrix $\rho$ (\ref{U19}) can be carried
out similarly to deformation of density matrix (density
pro-matrix) in Quantum Mechanics at Planck's scale (see
\cite{r14},\cite{r15}). Proceeding with this analogy density
matrix $\rho$ in (\ref{U19})
 should be changed by $\rho(\tau)$, where $\tau$ is a parameter of deformation.
Deformed density matrix  must fulfill the condition
$\rho(\tau)\approx\rho$ when $T\ll T_{p}$. By analogy with
\cite{r14},\cite{r15}, only probabilities $\omega_{n}$ are subject
of deformation in (\ref{U19}), changing by $\omega_{n}(\tau)$ and
correspondingly deformed statistical density matrix is
\begin{equation}\label{U20}
\rho(\tau)=\sum_{n}\omega_{n}(\tau)|\varphi_{n}><\varphi_{n}|.
\end{equation}

 This approach in our
opinion could give us the possibility to obtain Deformed Canonical
Distribution as well as a rigorous proof of thermodynamical
general uncertainty relations.

An detailed analysis of deformation in Statistical Mechanics is an
object of future investigations.

\end{document}